\documentclass[english]{aastex}
\usepackage[T1]{fontenc}
\usepackage{color}
\usepackage{amsbsy}
\usepackage{graphicx}

\makeatletter
\slugcomment{}
\shorttitle{}
\shortauthors{}

\makeatother

\usepackage{babel}
\begin{document}

\title{Secondary Rayleigh-Taylor type Instabilities in the Reconnection
Exhaust Jet as a Mechanism for Supra-Arcade Downflows }

\author{L.-J. Guo\altaffilmark{1,2,3,4}}

\email{lijiag@princeton.edu}

\author{A. Bhattacharjee\altaffilmark{1,2,3,4}}

\email{amitava@princeton.edu}

\author{Y.-M. Huang\altaffilmark{1,2,3,4}}

\email{yiminh@princeton.edu}

\and{}

\author{D. E. Innes\altaffilmark{4,5}}

\email{\textcolor{black}{innes@mps.mpg.de}}

\altaffiltext{1}{Space Science Center, University of New Hampshire, Durham, NH
03824, U.S.A.}

\altaffiltext{2}{Department of Astrophysical Sciences and Princeton Plasma Physics
Laboratory, Princeton University, Princeton, NJ 08540, U.S.A.}

\altaffiltext{3}{Max Planck/Princeton Center for Plasma Physics, Princeton, NJ
08540, U.S.A.}

\altaffiltext{4}{Max-Planck Institute for Solar System Research, Göttingen 37077,
Germany}
\begin{abstract}
Supra-arcade downflows (hereafter referred to as SADs) are low-emission,
elongated, finger-like features usually observed in active-region
coronae above post-eruption flare arcades. Observations exhibit downward
moving SADs intertwined with bright upward moving spikes. Whereas
SADs are dark voids, spikes are brighter, denser structures. Although
SADs have been observed for decades, the mechanism of formation of
SADs remains an open issue. In our three-dimensional resistive magnetohydrodynamic
simulations, we demonstrate that secondary Rayleigh-Taylor type instabilities
develop in the downstream region of a reconnecting current sheet.
The instability results in the formation of low-density coherent structures
that resemble SADs, and high-density structures that appear to be
spike-like. Comparison between the simulation results and observations
suggests that secondary Rayleigh-Taylor type instabilities in the
exhaust of reconnecting current sheets provide a plausible mechanism
for observed SADs and spikes.
\end{abstract}


\section{Introduction}

Supra-arcade downflows (hereafter referred to as SADs) (also known
as tadpoles due to their wavy appearance) are low-emission, elongated
features usually observed in active-region coronae above post-eruption
flare arcades (\citealt{mckenzi99,mckenzie00}). SADs ar\textcolor{black}{e
usually observed in }the extreme-ultraviolet (EUV)\textcolor{black}{{}
and X-ray filter images that detect plasma in the temperature range
$10^{6.8}-10^{7.3}$ Kelvin, and they have a typical life time of
a few minutes. By} using the filter ratio method (\citealt{hara92})
to deduce the temperature with the data from the Soft X-ray Telescope
(SXT), \citet{mckenzi99} showed that SADs are low-density ($<10^{9}cm^{-3}$),
high-temperature ($\sim10^{7}K$) structures. This result is supported
by \citet{savage12} as well as the SUMER spectroscopic analysis conducted
by \citet{innes03a}. \citet{asai04} found that the occurrences of
SADs are highly correlated with nonthermal bursts in microwave and
hard X-ray (HX\textcolor{black}{R), suggesting that the formation
of SADs involves magnetic reconnection or consequent outflows. \citet{innes03b}
reported high Doppler-shifted Fe XXI line profiles at the edges of
SADs, corresponding to line-of-sight velocity up to 1000 km/s. However,
the source of} the observed high-velocity is not clearly established
by these studies. More recently, \citet{savage11} conducted a statistical
study and found that the average velocity of most SADs is around 150km/s,
which is a fraction of the typical Alfvén speed ($\sim$1000 km/s)
of the supra-arcade plasma in corona. Furthermore, \citet{mckenzie13}
performed local correlation tracking (LCT) on sequences of EUV images
and found that vortices existed at the regions where SADs were observed.

It is important not to confuse SADs with plasmoids or magnetic islands.
Observationally, SADs are density depletion regions (\citealt{savage12,innes03a}),
whereas plasmoids are usually density-enhanced structures (\citealt{lin05,liu10}).
Plasmoids are observed edge-on as bright blobs moving along the post-CME
current sheet, whereas SADs are most clearly visible when observing
the current sheet\textcolor{black}{{} and the underlying arcade face-on.
The SADs and spikes are seen intermittently in the direction perpendicular
to the underlying arcades and appear to be flute-like (i.e. $\mathbf{k}\cdot\mathbf{B}\simeq0$,
where $\mathbf{k}$ is the wave number and $\mathbf{B}$ is the magnetic
field). This distinction is clearly illustrated in \citet{asai04}
and \citet{savage12}.}

Although SA\textcolor{black}{Ds have been the subject of significant
theoretical research during the past decade, the physical mechanisms
that drive the formation of finger-like SADs and spikes remain under
debate. There have been some simulations attempting to reproduce the
observational features of SADs in the literature. The} ``patchy reconnection''
model (e.g., \citealt{linton06,linton09}) uses \textcolor{black}{spatially
localized anomalous resistivity} intermittently over time along the
current sheet layer to trigger\textcolor{black}{{} intermittent reconnection.
}The reconnected magnetic field lines then cause intermittent disturbances
in the current sheet as they contract toward the downstream region,
and the flux tubes that emerge, with tear drop-like cross-sections,
have been interpreted as SAD\textcolor{black}{s. To test this idea,
\citet{scott13} attempt to reproduce SADs in simulation as the wakes
caused by reconnected flux tubes moving at a high speed. On the }other
hand, the SAD model developed by \citet{Costa09}, \citet{Maglione11},
and \citet{Cecere2012} assumes multiple reconnection sites in which
the SADs are a consequence of shocks and rarefactions bouncing back
and forth in magnetic structures. In these studies, magnetic reconnection
is not directly simulated; instead, reconnection ejections are modeled
with localized pressure enhancements in the initial condition. Recently,
\citet{cassak13} proposed that SADs are flow channels carved by low-density,
sunward-directed reconnection jets in high-density underlying arcades.
In this scenario, reconnection is continuous in time so that the SADs
are not filled in from behind as they would be if they were caused
by isolated descending flux tube\textcolor{black}{s; however, reconnection
has to be spatially localized to keep the outflow jet collimated. }

The studies mentioned above have a common feature that reconnection
has to be at least spatially localized, while in some scenarios reconnection
has to be temporally localized as well. In this Letter, we show that
the finger-like SADs can arise as a result of secondary Rayleigh-Taylor
type instabilities in the downstream region of reconnection in a post-eruption
current sheet. This physical mechanism was first suggested in \citet{asai04}
and explored partially by \citet{tandokoro05} by means of magnetohydrodynamic
(MHD) simulations primarily in the context of the Earth's magnetotail.
In this work we carry out two simulations, one with uniform resistivity
and the other with spatially localized anomalous resistivity. We find
that Rayleigh-Taylor type instabilities arise in both simulations.
Dynamic features of the instabilities exhibit good agreement with
observations of SADs, and the emulated synthetic emission count rate
from simulations also shows qualitative similarities with extreme
ultra-violet (EUV) images obtained from the Atmospheric Imaging Assembly
(AIA) on board the Solar Dynamics Observatory (SDO). \textcolor{black}{For
a more comprehensive discussion of observations of SADs, readers are
referred to our recent paper, \citet{innes14}. Inte}restingly, the
uniform resistivity simulation appears to be in better agreement with
observations.

\section{Simulation}

In this section, we introduce the setup of our MHD simulation for
studies of SADs, the simulation results, and examine their consistency
with observations.

\subsection{Simulation setup}

Our numerical model solves the following normalized three-dimensional
MHD equations 

\begin{equation}
\partial_{t}\rho=-\nabla\cdot\left(\rho\mathbf{v}\right),\label{eq:1}
\end{equation}
\begin{equation}
\partial_{t}\left(\rho\mathbf{v}\right)=-\nabla\cdot\left(\rho\mathbf{vv}\right)-\nabla p+\mathbf{J}\times\mathbf{B}+\mu\nabla^{2}(\rho\mathbf{v}),\label{eq:2}
\end{equation}
\begin{equation}
\partial_{t}p=-\nabla\cdot(p\mathbf{v})-(\gamma-1)p\nabla\cdot\mathbf{v}+(\gamma-1)\eta J^{2},\label{eq:3}
\end{equation}
\begin{equation}
\partial_{t}\boldsymbol{B}=-\nabla\times\left(-\mathbf{v}\times\boldsymbol{B}+\eta\mathbf{J}\right).\label{eq:4}
\end{equation}
Here $\rho$ is the plasma density, $\mathbf{v}$ is the plasmas velocity,
$\mathbf{B}$ is the magnetic field, $\mathbf{J}=\nabla\times\mathbf{B}$
is the electric current density, $p$ is the plasma thermal pressure,
$\mu$ is the plasma viscosity, and $\eta$ is the plasma resistivity.
Our model includes ohmic heating, but exclude viscous heating, heat
conduction and radiation cooling. Gravity is not included.

We consider a triply periodic system in the domain $-L_{x}\le x\le L_{x}$
, $-L_{y}\le y\le L_{y}$, $-L_{z}\le z\le L_{z}$. However, simulations
are carried out in the region $0\le x\le L_{x}$ and $0\le z\le L_{z}$,
and solutions in the remaining part of the domain are inferred by
symmetry. The initial magnetic field is the Harris double current
layer \citep{tandokoro05} defined as:
\[
B_{x}=\left\{ \begin{array}{cc}
B_{0}\tanh(z/a) & |z|\le L_{z}/2\\
-B_{0}\tanh((z-L_{z})/a) & z>L_{z}/2
\end{array}\right.,
\]
where $a$ is the Harris current sheet width, and $B_{0}$ is the
asymptotic magnetic field strength in the upstream region. The plasma
thermal pressure is calculated by the force balance condition:

\[
p+\frac{B_{x}^{2}}{2}=\frac{B_{0}^{2}}{2},
\]
and the density profile is determined by the ideal gas law $p=2\rho T_{0}$,
where $T_{0}$ is the constant initial temperature and the factor
$2$ is due to contributions from both electrons and ions. In normalized
units, we set $T_{0}=0.125$, $B_{0}=1$, $a=1$, and $\mu=5\times10^{-4}$.
The density $\rho=1$, the plasma beta $\beta\equiv2p/B_{0}^{2}=0.5$,
and the Alfven speed $V_{A}=1$ in the asymptotic region. The initial
density profile is seeded with a random noise o\textcolor{black}{f
magnitude $3\times10^{-2}$ to facilitate the instability.} The dimensions
of the simulation box are $L_{x}=150$, $L_{y}=5$, $L_{z}=40$, with
grids $n_{x}=800$, $n_{y}=150$, $n_{z}=400$, respectively. The
$+x$ direction is interpreted as the sunward direction. With this
basic setting, we have carried out simulations with uniform, as well
as with spatially localized anomalous resistivity.

\subsection{Uniform resistivity run}

In the first run, we use a uniform resistivity $\eta=3\times10^{-3}$
for the whole simulation box. To start the reconnection, we add an
initial perturbation to the magnetic field in the $x$ and $z$ directions: 

\[
\delta B_{x}=-0.025B_{0}\frac{L_{x}}{L_{z}}\cos(\frac{\pi x}{L_{x}})\sin(\frac{\pi z}{L_{z}}),
\]
\[
\delta B_{z}=0.025B_{0}\sin(\frac{\pi x}{L_{x}})\cos(\frac{\pi z}{L_{z}}).
\]
Subsequently, magnetic reconnection occurs along an extended Sweet-Parker
current sheet.\textcolor{black}{{} Reconnected magnetic field lines
form magnetic arcades in the downstream region, where} plasma ejected
by the reconnection outflows accumulate and form a high-density region.
\textcolor{black}{An interface is seen to form between lower density
reconnection outflows and the higher density plasma above the top
of the arcades. Eventually, the interface becomes wavy, and the small
wavy fluctuations develop into finger-like structures. Panel (a) of
Figure}\textcolor{blue}{{} \ref{fig2}}\textcolor{black}{{} shows the
lower part of the extended Sweet-Parker current sheet and the cusp
region at the top of flare arcades, where the finger-like structures
form.} Panel (b) of Figure \ref{fig2} shows a 2D slice of the density
profile in the $x\mbox{-}y$ plane at $z=0.1$, panel (c) shows the
temperature profile and panel (d) shows the expected count rate (DN/s/pixel)
in the AIA $131\ \mathring{A}$ channel calculated from the simulation
data at the same plane.\textcolor{black}{{} }The emission count rate
is calculated according to the formula $CR=\int n^{2}f(T)dl\ DN/s\cdot pixel$,
where $f(T)$ is the AIA $131\ \mathring{A}$ response function (\citealt{lemen12}),
where $n$ is the electron number density, $T$ is the temperature
and $dl$ is the line element along the line of sight. To use the
response function, the plasma density and temperature have to be converted
to dimensional units. Here the density is converted by assuming that
unit density in simulation equals $10^{9}\mbox{cm}^{-3}$. The temperature
is converted by assuming that the initial normalized Alfven speed
$V_{A}=1$ in the lobe corresponds to $V_{A}=1000\mbox{km/s}$, which
gives the initial temperature $T_{real}=\frac{m_{p}V_{A}^{2}}{k}T_{code}=\frac{1.67\times10^{-27}\mbox{kg}\times(10^{6}\mbox{m/s})^{2}}{1.38\times10^{-23}\mbox{m}^{2}\mbox{kg}\mbox{s}^{-2}\mbox{K}^{-1}}\times0.125\simeq1.5\times10^{7}\mbox{K}$
for our simulation. Likewise, the initial normalized magnetic field
in the lobe ($B_{code}=1$) corresponds to $B_{real}\simeq14\mbox{\ Gauss}$,
which is a reasonable value for the coronal magnetic field. 

The finger-like structures are caused by plasma instabilities in the
exhaust region of a reconnecting current sheet. As can be seen from
Figure \ref{fig2}, the instabilities take place at the interface
between lighter reconnection outflows and denser plasma (piled-up
density in front of reconnection outflows). Because the reconnection
outflows push the relatively stationary plasma ahead, the deceleration
existing between lighter and denser plasma plays a role that is equivalent
to gravity in the Rayleigh-Taylor instability. Furthermore, the magnetic
field lines in the arcade are highly bent in the downstream region
(Figure \ref{fig2}), producing unfavorable curvature that makes the
system potentially unstable to the ballooning instability (cf. \citealt{bhattacharjee98}).
While we have not carried out a detailed linear stability analysis
and leave this to future work, we propose that the finger-like SADs
in the downstream region are results of the nonlinear evolution of
instabilities of the Rayleigh-Taylor/ballooning type.

The typical speed of the SADs (speed of the tip motion) in this run
is $\sim0.05V_{A}$, which is comparable to the observed value. The
instabilities also induce transverse motion of SADs. Figure \ref{fig3}
shows the flow pattern on the $x-y$ plane, overplotted on the density
profile shown in color. Downward moving tadpoles and upward moving
spikes can be seen in Figure \ref{fig3}, as well as vortices. The
plasma in the cusp region/supra-arcade fan is quite turbulent due
to secondary instabilities. The observations of eddies co-existing
with SADs have been reported by \citet{mckenzie13}.\textbf{ }

\textcolor{black}{In observations, the clusters of SADs happen repeatedly
over the course of the supra-arcade evolution. We found a similar
behavior in our simulation. From }the four panels of Figure \ref{fig2},
we see that there are two clusters of finger-like structures, one
at $x\simeq115$ and the other at $x\simeq130$. Throughout the whole
simulation, there are four clusters of finger-like structures. \textcolor{black}{Because
the resistivity is uniform in space and constant over time in this
run, our simulation suggests that the intermittent formation of finger-like
SADs does not necessarily require intermittent, locally enhanced resistivity,
but can be attributed to ideal instabilities in the downstream region. }

\subsection{Anomalous resistivity run}

To test how the reconnection mechanism may affect the instabilities
in the downstream region, we have carried out a second run with Petschek-type
reconnection triggered by introducing a locally enhanced anomalous
resistivity $\eta=\eta_{0}\exp(-x^{2}-z^{2})$, with $\eta_{0}=3\times10^{-3}$.
The Petschek-type reconnection soon creates a shock-like front propagating
along the $+x$ direction, with the $B_{z}$ component and the plasma
density $\rho$ piled up on the downstream side of the front.\textcolor{black}{{}
This propagating front later develops wavy structures, which subsequently
grow and became elongated in the $x$ direction, }as shown in Figure
\ref{fig4}. Panel (a) of Figure \ref{fig4} shows a projected view
of magnetic field lines, and a slice of the $B_{z}$ profile at $z=0.1$.
Panel (b) of Figure \ref{fig4} shows a 2D slice of the density profile
on the $x-y$ plane at $z=0.1$. Panel (c) shows the temperature profile,
and panel (d) shows the AIA $131\ \mathring{A}$ emission count rate
calculated from the simulation data on the same plane. 

The finger-like structures in Figure \ref{fig4} consist of two parts.
The ``tadpoles'' move in the $+x$ (sunward) direction and have
low density, high temperature, and weak emission, and the ``spikes''
are fingers developing toward the $-x$ (anti-sunward) direction with
higher density, lower temperature, and brighter emission.\textcolor{red}{{}
}\textcolor{black}{At a later stage, secondary instabilities (e.g.
the Kelvin-Helmholtz instability) may happen at the interface between
tadpoles and spikes and eventually cause mixing of tadpoles and spikes.
Figure \ref{fig5} shows the line of sight ($z$-direction) averaged
emission count rate at four different times. The top panel shows tadpole-like
structures at $t=120$. In the second and the third panels ($t=140$
and $160$), the tadpoles get thinner as they descend further. Finally,
in the fourth panel ($t=180$), the tadpoles merge into the brighter
surroundings. }

Note that although a locally enhanced anomalous resistivity is employed
in this run, the localized resistivity is uniform in the $y$ direction
(along the current sheet layer) and constant over time. Therefore,
the finger-like SADs are caused by instabilities and not a direct
result of anomalous resistivity. Furthermore, the SADs in this run
are qualitatively similar to the ones we obtained with the uniform
resistivity. This suggests that the reconnection mechanism in the
upstream region does not directly affect instabilities in the downstream
region. Nevertheless, there are some differences regarding where and
when the SADs take place and the visual propagation speed of SADs,
which we will discuss in the next section.

\section{Summary and Conclusions }

The mechanism causing the formation of SADs has been an open question
since their first discovery. Existing simulations of SADs depend on
intermittently and locally induced reconnection events to reproduce
finger-like SADs along current sheet layers. In addition, questions
such as why the speeds of SADs are much slower than the Alfven speed
\textcolor{black}{and how the density depletion regions within SADs
manage to balance the thermal pressure from the surrounding plasmas
for a few minutes remain unsolved. In this} paper, we describe simulations
to test the idea that SADs might be the result of essentially ideal
plasma instabilities in the downstream region of a reconnection site.
We implement resistive MHD models with both uniform and spatially
localized anomalous resistivity. Finger-like structures are generated
in both runs with different dynamic behaviors. The comparisons between
our simulations and observations of SADs can be summarized as follows:

\textcolor{black}{(1) Appearance. The downward developing parts (tadpoles)
of the finger-like structures are characterized by low density, high
temperature, and low EUV emission count rate (shown in Figure \ref{fig2}
and Figure \ref{fig4}). The upward developing parts (spikes) are
characterized by higher density, relatively low temperature and higher
EUV emission count rate. The features of the downward developing parts
in the simulations are very similar to the observational features
of SADs, while the upward developing parts resemble bright spikes
observed among SADs. The relatively higher temperature in the low-density
plasma of tadpoles keeps them in approximate force balance with surrounding
plasma and spikes. Therefore, our simulations are able to successfully
reproduce the observational features of SADs and explain why elongated
SADs structures can exist for a few minutes. In addition, both simulations
and observations show that SADs get thinner as they penetrate deeper
into the bright fans. Our simulations suggest that this behavior may
be due to shear flow instabilities between upward-moving, high density
plasma in spikes and downward-moving, low density plasma in tadpoles,
which eventually result in the merger of tadpoles with the surrounding
plasma. }

\textcolor{black}{(2) The timeline and the location. Observations
show that after the initial eruption of a CME, bright material starts
to accumulate above existing coronal loops. After a while, SADs start
to show up as dark flows penetrating the bright material above the
corona loops and bright spikes start to appear between the SADs. The
whole event can last for a few hours, during which SADs occur repeatedly.
The observations suggest the following scenario. After a CME eruption,
a post-eruption reconnection site is formed; as magnetic reconnection
proceeds, reconnection outflows develop in both the sunward and anti-sunward
directions. The sunward moving outflows stack over the existing magnetic
arcades in the corona and heat up the surrounding plasma. As a result,
bright fans appear above the arcades. Later on, a cluster of SADs
develop at the top of the bright fans due to instabilities and the
low density jet plasma descends while the bright spikes move upward.
In the following several hours, clusters of SADs occur repeatedly.
The magnetic configuration and the timeline of events described here
are in good agreement with the ones in the simulation with uniform
resistivity. }

\textcolor{black}{(3) Dynamic characteristics. In most events, the
SADs move sunward at speeds much lower than the Alfven speed \citep{savage11},
typically 15\% of the Alfvén speed. In the run with uniform resistivity,
the instabilities occur at the top of the magnetic arcades where reconnection
outflows have decelerated almost to a standstill. The velocity of
SADs are seen in the simulations is approximately 5\% of the Alfvén
speed, which is in the range of the observed values. However, in the
run with anomalous resistivity, the instabilities develops at the
shock-like front which propagates sun-ward at the Alfvénic speed,
which is not in good agreement with observations. From this perspective,
the MHD model with uniform resistivity agrees better with observations
than the one with anomalous resistivity.}

Our results indicate that the exhaust region of magnetic reconnection
can become unstable to Rayleigh-Taylor type instabilities, and the
resulting finger-like structures exhibit features qualitatively similar
to observations of SADs. Most importantly, we have shown that SADs
can arise without reconnection being patchy, i.e. localized in either
space or time (or both). In fact, the reconnection site in both runs
of this study does not develop apparent 3D (i.e. $y$-dependent) structures,
and reconnection continues at a constant rate throughout the simulations.
Of course, in reality the reconnection site is likely to exhibit 3D
effects, either due to the local onset of reconnection or 3D instabilities
in the current sheet. It will be interesting to study how the 3D nature
of the reconnection site affects the structures in the exhaust region.\textcolor{black}{{}
Our simulations can also account for the observed transverse motion
of SADs (e.g. merging and splitting of SADs and vortices around SADs).
We also note that in the run with uniform resistivity, SADs formed
at the top of fan spikes as seen in recent AIA observations (\citealt{innes14}).
The transverse motion of plasma may be due to shearing instabilities
between tadpoles and spikes. This kind of motion could make the local
region quite turbulent, as shown in Figure \ref{fig3}. }

In conclusion, our results suggest that Rayleigh-Taylor type instabilities
in the downstream region of a reconnecting current sheet provide a
plausible mechanism for the formation of SADs. Between the two models
of resistivity employed in our simulations, the uniform resistivity
appears to agree better with observations. Rayleigh-Taylor type instabilities
have also been observed in recent fully kinetic particle-in-cell simulations
(\citealt{Vapirev13}). Further study of the instabilities with different
underlying models and detailed comparison with observations should
shed new light on the nature of SADs, as well as what can be learned
about the structure of the reconnection site from the appearance of
SADs.

\acknowledgements{}

This work was supported \textcolor{black}{by the Department of Energy,
Grant No. DE-FG02-07ER46372, under the auspice of the Center for Integrated
Computation and Analysis of Reconnection and Turbulence (CICART),
the National Science Foundation, Grant No. PHY-0215581 (PFC: Center
for Magnetic Self-Organization in Laboratory and Astrophysical Plasmas),
NASA Grant Nos. NNX09AJ86G and NNX10AC04G, and NSF Grant Nos. ATM-0802727,
ATM-090315 and AGS-0962698. We acknowledge beneficial conversations
with Dr. Nicholas Murphy.}

\clearpage{}

\includegraphics[scale=0.5]{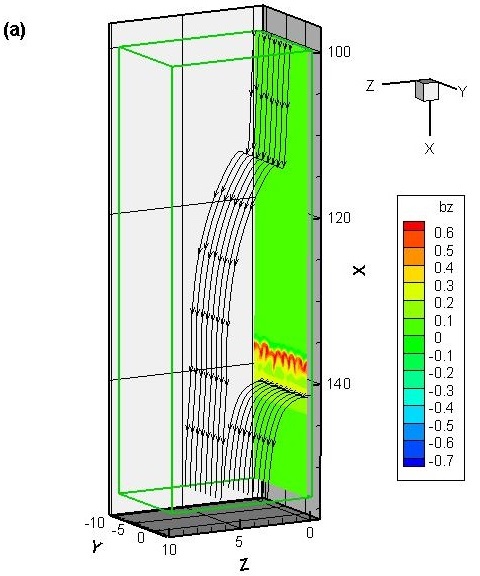} \includegraphics[scale=0.3]{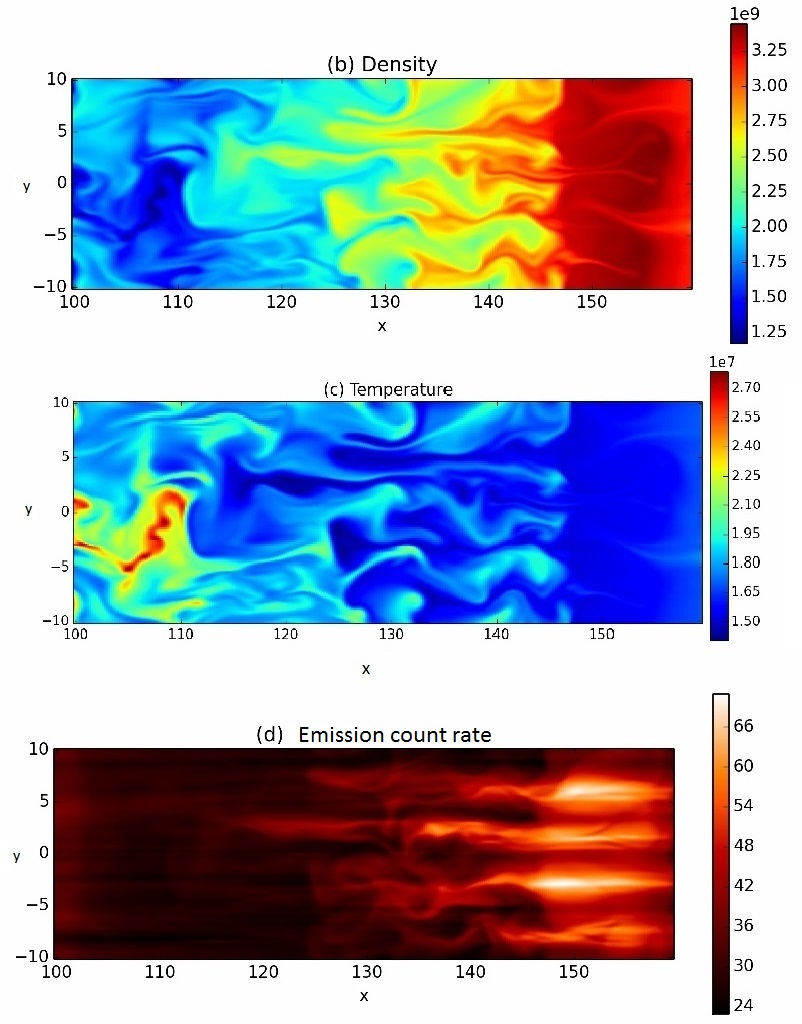}

\figcaption{\label{fig2} Panel (a) shows the magnetic field line configuration
and a $x-y$ slice of $B_{z}$ profile at $z=0.1$ from MHD model
with uniform resistivity, where the color codes represent the magnitude
of the $B_{z}$ component (unit: Ga\textcolor{black}{uss). The z-axis
in panel (a) is stretched by two times for a better projection. Pa}nel
(b) shows the density profile (unit : $/cm^{3}$), panel (c) shows
the temperature profile (unit: Kelvin), and panel (d) shows the synthetic
AIA 131 Å emission count rate on the same $x-y$ plane. }

\includegraphics[scale=0.45]{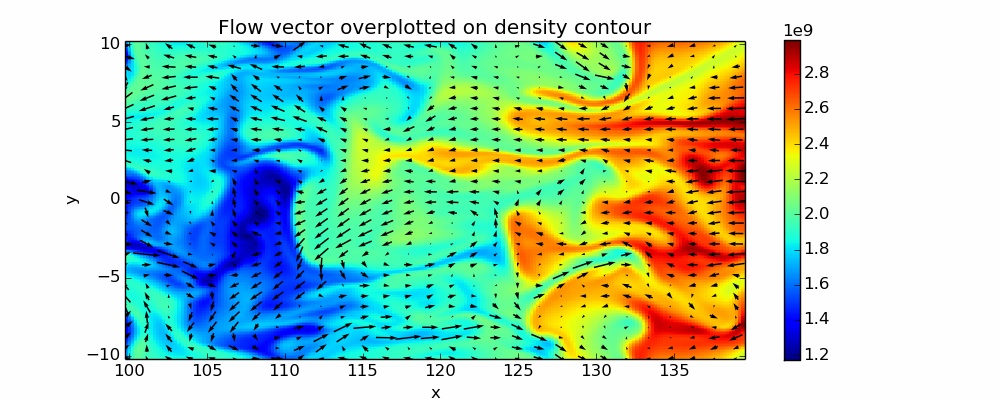}

\figcaption{\label{fig3} Flow pattern overplotted on a frame of the density
profile. }

\includegraphics[scale=0.38]{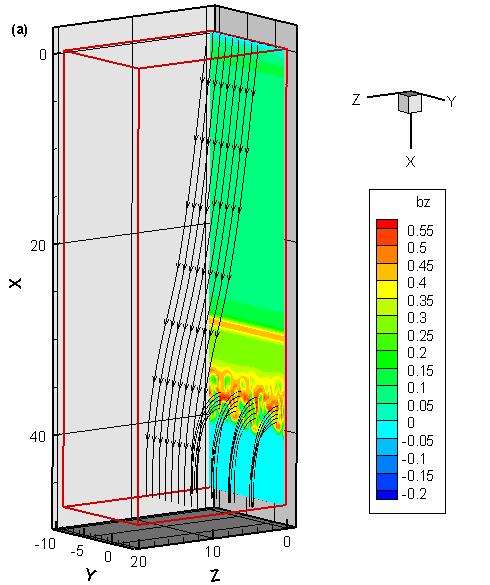} \includegraphics[scale=0.25]{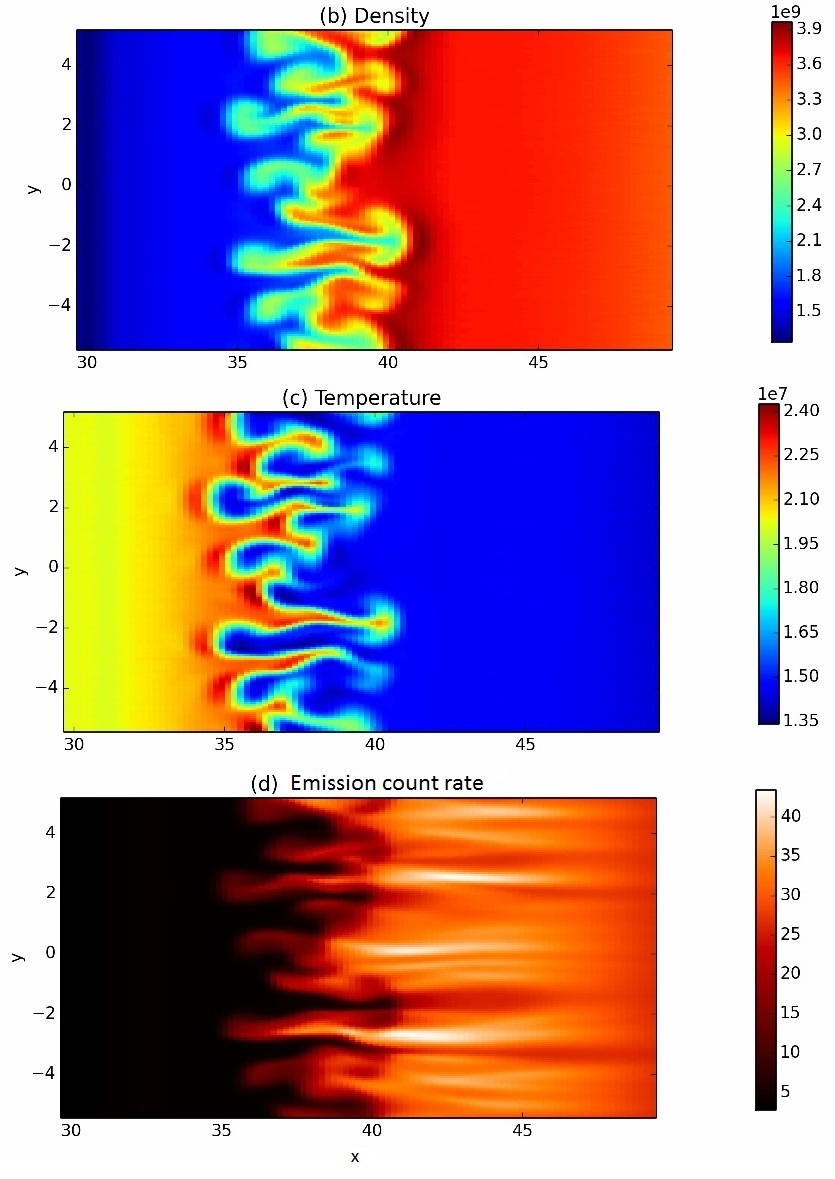}

\figcaption{\label{fig4} Panel (a) shows the magnetic field line configuration
and a $x-y$ slice of $B_{z}$ profile for MHD model with anomalous
resistivity. The 2D $x-y$ slice is placed at $z=0.1$, where the
color codes represent the magnitude of the $B_{z}$ component (unit:
Gauss). Panel (b) shows the density profile (unit: $/cm^{3}$), panel
(c) shows the temperature profile (unit: Kelvin), and panel (d) shows
the synthetic AIA $131\mathring{A}$ emission count rate on the same
$x-y$ plane.}

\includegraphics[scale=0.4]{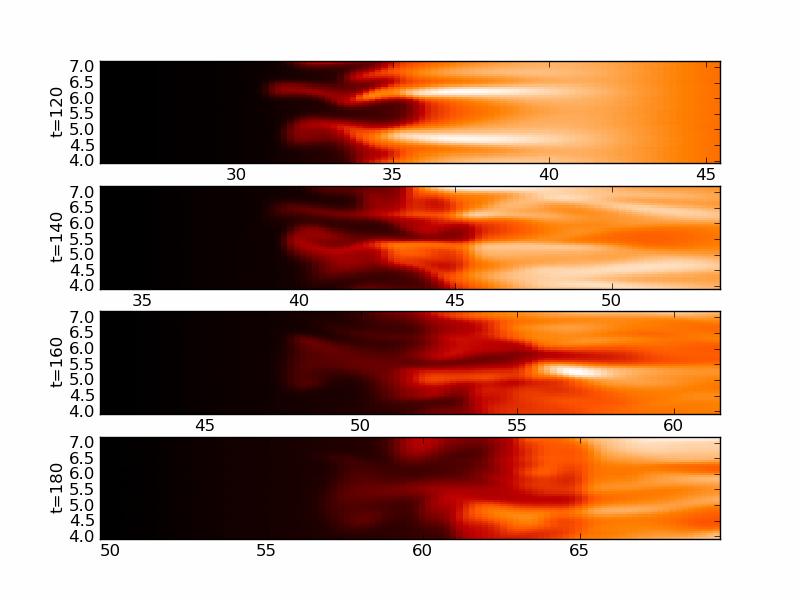}

\figcaption{\label{fig5} Line-of-sight averaged emission count rate featuring
one SAD event developing along the $x$-direction at four different
times. }
\end{document}